\documentclass[12pt,superscriptaddress,aps,prd,preprint]{revtex4}
\usepackage{amsmath}
\usepackage{amssymb}
\makeatletter
\newcommand{\bea}{\begin{eqnarray}}
\newcommand{\eea}{\end{eqnarray}}

\newcommand{\pa}{\partial}

\begin{document}
\title{The G\"{o}del solution in the modified gravity}
\author{C. Furtado}
\affiliation{Departamento de F\'{\i}sica, Universidade Federal da Para\'{\i}ba\\
Caixa Postal 5008, 58051-970, Jo\~ao Pessoa, Para\'{\i}ba, Brazil}
\email{furtado,jroberto,petrov,alesandro@fisica.ufpb.br}
\author{T. Mariz}
\affiliation{Instituto de F\'{\i}sica, Universidade Federal de Alagoas\\
CEP 57072-970, Macei\'{o}, Alagoas, Brazil}
\email{tmariz@if.ufal.br}
\author{J. R. Nascimento}
\affiliation{Departamento de F\'{\i}sica, Universidade Federal da Para\'{\i}ba\\
Caixa Postal 5008, 58051-970, Jo\~ao Pessoa, Para\'{\i}ba, Brazil}
\email{furtado,jroberto,petrov,alesandro@fisica.ufpb.br}
\author{A. Yu. Petrov}
\affiliation{Departamento de F\'{\i}sica, Universidade Federal da Para\'{\i}ba\\
Caixa Postal 5008, 58051-970, Jo\~ao Pessoa, Para\'{\i}ba, Brazil}
\email{furtado,jroberto,petrov,alesandro@fisica.ufpb.br}
\author{A. F. Santos}
\affiliation{Departamento de F\'{\i}sica, Universidade Federal da Para\'{\i}ba\\
Caixa Postal 5008, 58051-970, Jo\~ao Pessoa, Para\'{\i}ba, Brazil}
\email{furtado,jroberto,petrov,alesandro@fisica.ufpb.br}

\begin{abstract}
We consider the modified gravity whose action represents itself as a sum of the usual Einstein-Hilbert action and the gravitational Chern-Simons term and show that the G\"{o}del metric solves the modified equations of motion, thus proving that the closed timelike curves whose presence is characteristic for the G\"{o}del solution are not forbidden in the case of the Chern-Simons modified gravity as well.
\end{abstract}

\maketitle
The idea of modifications of the general relativity has a long history. The motivations for such modifications arise, first, from the perturbative studies of quantum gravity which have shown that the general relativity is non-renormalizable \cite{thooft}, second, from the astrophysical observations which have showed the accelerated expansion of the Universe \cite{Riess}. The most popular modifications are based on introduction of additive higher-derivative terms which are known to improve essentially renormalization properties of the theories \cite{stelle}.
At the same time, actually large scientific attention is attracted by the Lorentz and/or CPT breaking modifications of the gravity \cite{Kost}. The most popular of such modifications, with no doubts, is the additive gravitational Chern-Simons term \cite{JaPi} which represents itself as an example of the higher-derivative term. In \cite{JaPi} it was shown that this term introduces parity violation, i.e., each of two polarizations of gravitational waves travels with the speed of light, but with different intensity, however without Lorentz violation. Also, in \cite{JaPi} it was shown that the Schwarzschild solution is compatible with this modification of the gravity.  Some interesting results were obtained for this theory, for example, effects of this modification for bodies on orbits around the Earth were described in \cite{Smith}, the post-Newtonian expansion was studied in \cite{Alexander}, and some cosmological effects were analyzed in \cite{Lue,Alexander01}. A conserved, symmetric energy-momentum (pseudo)tensor for the Chern-Simons modified gravity was constructed in \cite{Guarrera}, with in \cite{Tekin} the conserved charges in this theory are discussed in details, in \cite{Jackiw01} it was shown that in this theory the Poincar\'e invariance holds, and in \cite{ours} the dynamical generation of this term via perturbative corrections was carried out. 

One of the most important issues related to the Chern-Simons modified gravity is the search for the solutions of the equations of motion for modified gravity. A discussion on this subject is presented in \cite{Grumiller}. In general, inclusion of the gravitational Chern-Simons term is characterized by the so-called external field $\theta$, and it was shown in \cite{Grumiller} that a wide class of solutions of the usual general relativity involving, in particular, spherically symmetric and axisymmetric solutions, persists to solve the modified gravity equations for specific forms of the $\theta$. However, some important metrics, such as Kerr metric, fail to solve the modified gravity equations and require essential modification \cite{Yunes,Konno}, with the modified metrics resolving the modified Einstein equations can be found only via the perturbative approach.

One of very important solutions in general relativity is the G\"{o}del metric \cite{Godel}, representing itself as a first cosmological solution with rotating matter. This solution is stationary, spatially homogeneous, possessing cylindrical symmetry, and its highly nontrivial property consists in breaking of causality implying in the possibility of the closed timelike curves (CTCs) in the G\"{o}del space, whereas, as it was conjectured by Hawking \cite{Hawk}, presence of CTCs is physically inconsistent. Furthermore, in \cite{Reb} this metric was generalized in cylindrical coordinates and the problem of causality was examined with more details, thus it turned out to be that one can distinguish three different classes of solutions. These solutions are characterized by the following possibilities:  (i)  there is no CTCs, (ii)  there is an infinite sequence of alternating causal and noncausal regions, and (iii)  there is only one noncausal region. In the paper \cite{Dabr} the quantities called superenergy and supermomentum which can be used as criteria of possibility of existence the CTCs were introduced. In \cite{vayona} the CTC solutions in the G\"{o}del space are discussed in the string context (for a review of different aspects of CTCs see also f.e. \cite{Lobo}). Another reasons for interest to the G\"{o}del solution consists in the fact that the G\"{o}del universe allows for nontrivially embedded black holes \cite{He}. Different aspects of the G\"{o}del solutions are discussed also in \cite{barrow}. In this paper we are interested in verifying whether the G\"{o}del solution holds in the Chern-Simons modified gravity.

We start our study with introducing the G\"{o}del metric which is written as \cite{Godel}
\bea
ds^2=a^2\Bigl[dt^2-dx^2+\frac{1}{2}e^{2x}dy^2-dz^2+2 e^x dt\,dy\Bigl],\label{godel}
\eea
where $a$ is a positive number. The non-zero Christoffel symbols corresponding to this metric look like 
\bea
\Gamma^0_{01}=1,\,\,\,\,\,\,\, \Gamma^0_{12}=\frac{1}{2}e^x,\,\,\,\,\,\,\,\Gamma^1_{02}=\frac{1}{2}e^x, \,\,\,\,\,\,\,\Gamma^1_{22}=\frac{1}{2}e^{2x},\,\,\,\,\,\,\,\Gamma^2_{01}=-e^{-x}.
\eea
The non-zero components of the Riemann tensor are
\bea
R_{0101}=-\frac{1}{2}a^2, \,\,\,R_{0112}=\frac{1}{2}a^2e^x, \,\,\, R_{0202}=-\frac{1}{4}a^2e^{2x}, \,\,\, R_{1212}=-\frac{3}{4}a^2e^{2x}.
\eea
The corresponding non-zero components of the Ricci tensor look like
\bea
 R_{00}=1, \,\,\,\,\,\,\,\, R_{02}=R_{20}=e^x, \,\,\,\,\,\,\,\, R_{22}=e^{2x}.
\eea
Finally, the Ricci scalar is
\bea
R=\frac{1}{a^2}.
\eea
 
It is easy to check that the G\"{o}del metric solves the Einstein equations
\bea
R_{\mu\nu}-\frac{1}{2}g_{\mu\nu}R=8\pi G\rho u_\mu u_\nu+\Lambda g_{\mu\nu},
\eea 
 where $u$ is a unit time-like vector whose explicit contravariant components look like $u^{\mu}=(\frac{1}{a},0,0,0)$  and the corresponding covariant components are $u_\mu=(a,0,ae^x,0)$.  
Indeed, let us for example consider the $(00)$ component of the Einstein equations, that is
\bea
R_{00}-\frac{1}{2}g_{00}R&=&8\pi G\rho u_0 u_0+\Lambda g_{00},
\eea
which leads to
\bea
\frac{1}{2}&=&8\pi G\rho a^2 +\Lambda a^2.
\eea
This equation is satisfied if
\bea
\Lambda=-\frac{1}{2a^2},\,\,\,\,\,\,\,\,\,\,\,\,8\pi G\rho=\frac{1}{a^2},
\eea
or, $\Lambda=-4\pi G\rho$. For the other components of the Einstein equations we find the same condition, i.e., the G\"{o}del metric solves the Einstein equations if and only if this condition is satisfied. This result was found in \cite{Godel}.

Now, let us modify the gravity action by adding the gravitational Chern-Simons term \cite{JaPi}.
The resulting Chern Simons modified gravity action looks like
\bea
\label{smod}
S=\frac{1}{16\pi G}\int d^4x\Bigl[\sqrt{-g}R+\frac{1}{4}\theta \,{^*}RR\Bigl] + S_{mat},
\eea
where $R$ is the scalar curvature, $S_{mat}$ is the matter action and ${^*}RR$ is the topological invariant called the Pontryagin term whose explicit definition is
\bea
{^*}RR\equiv {^*}{R^a}\,_b\,^{cd}R^b\,_{acd},
\eea
where $R^b\,_{acd}$ is the Riemann tensor and ${^*}{R^a}\,_b\,^{cd}$ is the dual Riemann tensor given by
\bea
{^*}{R^a}\,_b\,^{cd}=\frac{1}{2}\epsilon^{cdef}R^a\,_{bef}.
\eea
The function $\theta$ is an external scalar field. Alternatively, $\theta$ can be interpreted as a dynamical variable \cite{Grumiller,Yunes}, however, in this case the equations of motion acquire additional terms which makes the analysis more complicated \cite{Yunes}, therefore we consider $\theta$ as an external field henceforth. After integration by parts, introducing $v_{\mu}=\pa_{\mu}\theta$, we get
\bea
\frac{1}{4}\int d^4x \theta{^*}RR=-\int d^4x v_{\mu}\epsilon^{\mu\alpha\beta\gamma}(\frac{1}{2}\Gamma^{\sigma}_{\alpha\tau}\pa_{\beta}\Gamma^{\tau}_{\gamma\sigma}+\frac{1}{3}
\Gamma^{\sigma}_{\alpha\tau}\Gamma^{\tau}_{\beta\eta}\Gamma^{\eta}_{\gamma\sigma}),
\eea
that is, the usual gravitational Chern-Simons term.
Varying the action (\ref{smod}) with respect to the metric, we obtain the modified Einstein equations
\bea
 R^{\mu\nu}-\frac{1}{2}g^{\mu\nu}R+C^{\mu\nu}=8\pi G\rho u^\mu u^\nu, \label{einst}
\eea
where $C_{\mu\nu}$ is the Cotton tensor arising due to the varying of the additive Chern-Simons term, i.e. (cf. \cite{JaPi}). 
\bea
\delta \frac{1}{4}\int d^4x \theta {}^*RR=\int\sqrt{-g}C^{\mu\nu}\delta g_{\mu\nu}.
\eea
The explicit form of the Cotton tensor is \cite{DJPi}
\bea
C^{\mu\nu}=-\frac{1}{2\sqrt{-g}}\Bigl[v_\sigma \epsilon^{\sigma\mu\alpha\beta}D_\alpha R^\nu_\beta+\frac{1}{2}v_{\sigma\tau}\epsilon^{\sigma\nu\alpha\beta}R^{\tau\mu}\,_{\alpha\beta}\Bigl] \,+\, (\mu\longleftrightarrow\nu), 
\eea
with $v_\sigma\equiv\partial_\sigma \theta$, $v_{\sigma\tau}\equiv D_\sigma v_\tau$.  Taking the covariant divergence of this equation we find
\bea
D_\mu C^{\mu\nu}=\frac{1}{8\sqrt{-g}}v^{\nu}{^*}RR.
\eea
Using the Bianchi identity, $D_\mu G^{\mu\nu}=0$, and suggesting that the mattter terms are diffeomorphism invariant, i.e. $D_\mu T^{\mu\nu}=0$, with $T^{\mu\nu}=\rho u^\mu u^\nu$, one finds that the solution of the equation (\ref{einst}) requires a consistency condition,
\bea
{^*}RR=0.
\eea
This consistency condition implies that the diffeomorphism symmetry breaking is suppressed on-shell (for more details see \cite{JaPi, Jackiw01}).

As the G\"{o}del metric solves the usual Einstein equations, it can solve the modified ones if and only if the Cotton tensor vanishes for such a metric. Let us verify whether it is the case.

The nontrivial components of the Cotton tensor can be explicitly found and look like
\bea
C^{00}&=&\frac{1}{\sqrt{-g}}\frac{e^x}{a^2}\,\Bigl[2v_3+v_{31}\Bigl],\nonumber\\
C^{01}&=&\frac{1}{2\sqrt{-g}}\frac{1}{a^2}\,\Bigl[v_{32}-e^x v_{30}\Bigl],\nonumber\\
C^{02}&=&-\frac{1}{2\sqrt{-g}}\frac{1}{a^2}\,\Bigl[2v_3+v_{31}\Bigl],\nonumber\\
C^{03}&=&\frac{1}{2\sqrt{-g}}\frac{1}{a^2}\,\Bigl[v_2-2e^x v_0-e^x v_{01}\Bigl],\nonumber\\
C^{11}&=&\frac{1}{\sqrt{-g}}\frac{e^x}{2a^2}\,v_3,\nonumber\\
C^{13}&=&-\frac{1}{2\sqrt{-g}}\frac{1}{2a^2}\,\Bigl[e^x v_1+v_{20}-2e^x v_{00}+v_{02}\Bigl],\nonumber\\
C^{22}&=&\frac{1}{\sqrt{-g}}\frac{e^{-x}}{a^2}\,v_3,\nonumber\\
C^{23}&=&\frac{1}{2\sqrt{-g}}\frac{1}{2a^2}\,\Bigl[-2e^{-x} v_2+2v_0+v_{10}+v_{01}\Bigl].
\eea
The remaining components, that is, $C^{12}$ and $C^{33}$, are identically equal to zero. Using the definition of $v_\sigma$ and writing $v_{\sigma\tau}=\partial_\sigma\partial_\tau\theta-\Gamma^\lambda_{\sigma\tau}v_\lambda$, we can express the non-zero components of the Cotton tensor as
\bea
C^{00}&=&\frac{1}{\sqrt{-g}}\frac{e^x}{a^2}\,\Bigl[2\partial_3\theta+\partial_3\partial_1\theta\Bigl],\nonumber\\
C^{01}&=&\frac{1}{2\sqrt{-g}}\frac{1}{a^2}\,\Bigl[\partial_3\partial_2\theta-e^x \partial_3\partial_0\theta\Bigl],\nonumber\\
C^{02}&=&-\frac{1}{2\sqrt{-g}}\frac{1}{a^2}\,\Bigl[2\partial_3\theta+\partial_3\partial_1\theta\Bigl],\nonumber\\
C^{03}&=&\frac{1}{2\sqrt{-g}}\frac{1}{a^2}\,\Bigl[-e^x \partial_0\theta-e^x\partial_0\partial_1\theta \Bigl],\nonumber\\
C^{11}&=&\frac{1}{\sqrt{-g}}\frac{e^x}{2a^2}\,\partial_3\theta,\nonumber\\
C^{13}&=&-\frac{1}{2\sqrt{-g}}\frac{1}{2a^2}\,\Bigl[-2e^x \partial_0\partial_0\theta+\partial_0\partial_2\theta+\partial_2\partial_0\theta\Bigl],\nonumber\\
C^{22}&=&\frac{1}{\sqrt{-g}}\frac{e^{-x}}{a^2}\,\partial_3\theta,\nonumber\\
C^{23}&=&\frac{1}{2\sqrt{-g}}\frac{1}{2a^2}\,\Bigl[\partial_1\partial_0\theta+\partial_0\partial_1\theta\Bigl].
\eea

Now, analysing these results, we note that the G\"{o}del metric can solve the modified gravity equation only for the specific form of the external field $\theta$, that is,
\bea
\label{thf}
\theta=F(x,y),
\eea
which, in particular case, can be rewritten as $\theta=F(x)+xG(y)$, i.e., as a reminiscence of the structure of the $\theta$ used in \cite{Grumiller} for the spherically symmetric cases. Indeed, in this case all components of the Cotton tensor vanish, thus making the G\"{o}del metric to be a solution of the modified Einstein equations. Therefore, for this specific choice of the $\theta$ function, we find that the G\"{o}del metric is compatible with the Chern Simons modified gravity. As a consequence, the highly nontrivial property of the G\"{o}del metric, that is, the situation with the possibility for  existence of the CTCs in the Chern-Simons modified gravity in the case of the external field $\theta$ in the form (\ref{thf}) does not differ from the case of the usual Einstein gravity, and the discussion of \cite{Reb,Dabr} is applicable in the modified gravity case as well as in the usual case. 

{\bf Acknowledgements.} The work has been supported by Conselho Nacional de
Desenvolvimento Cient\'\i fico e Tecnol\'ogico (CNPq), by Coordena\c c\~ao de Aperfei\c coamento de Pessoal de N\'\i vel Superior (CAPES: AUX-PE-PROCAD 579/2008) and by CNPq/PRONEX/FAPESQ.


\begin{thebibliography}{99}
\bibitem{thooft} G. 't Hooft, M. Veltman, Ann. Inst. Henri Poincare, vol. XX, 69 (1974); M. Veltman, "Quantum theory of gravitation", in: Les Houches, Session XXVIII, North-Holland, 1976 -- Methods in Field Theory (eds. R. Balian, J. Zinn-Justin), p. 265-327.
\bibitem{Riess} A. Riess et al, Astron. J. 116, 1009 (1998), astro-ph/9805201.
\bibitem{stelle} K. Stelle, Phys. Rev. D {\bf 16}, 953 (1977).
\bibitem{Kost} V. A. Kostelecky, Phys. Rev. D {\bf 69}, 105009 (2004), hep-th/0312310.
\bibitem{JaPi} R. Jackiw, S.-Y. Pi, Phys. Rev. D {\bf 68}, 104012 (2003), gr-qc/0308071.
\bibitem{Smith} T.L. Smith, A. L. Erickcek, R. R. Caldwell, M. Kamionkowski, Phys. Rev. D {\bf 77}, 024015 (2008), astro-ph/0708.0001. 
\bibitem{Alexander} S. Alexander, N. Yunes, Phys. Rev. D {\bf 75}, 124022 (2007).
\bibitem{Lue} A. Lue, L. Wang, M. Kamionkowski, Phys. Rev. Lett. {\bf 83}, 1506 (1999), astro-ph/9812088.
\bibitem{Alexander01} S. Alexander, Phys. Rev. Lett. B {\bf 660}, 444 (2008), hep-th/0601034.
\bibitem{Guarrera} D. Guarrera, A. J. Hariton, Phys. Rev. D {\bf 76}, 044011 (2007), gr-qc/0702029. 
\bibitem{Tekin} B. Tekin, Phys. Rev. D {\bf 77}, 024005 (2008), arXiv: 0710.2528 [gr-qc].
\bibitem{Jackiw01} R. Jackiw, "Lorentz violation in a diffeomorphism-invariant theory", arXiv: 0709.2348 [hep-th].
\bibitem{ours} T. Mariz, J. R. Nascimento, A. Yu. Petrov, L. Y. Santos, A. J. da Silva, Phys. Lett. B661, 312 (2008), arXiv: 0708.3348 [hep-th];  M. Gomes, T. Mariz, J. R. Nascimento, A. Yu. Petrov, E. Passos, A. J. da Silva, Phys. Rev. D78, 025029 (2008), arXiv: 0805.4409 [hep-th].
\bibitem{Grumiller} D. Grumiller and N. Yunes, Phys. Rev. D {\bf 77}, 044015 (2008), arXiv: 0711.1868 [hep-th].
\bibitem{Yunes} N. Yunes, F. Pretorius, Phys. Rev. D {\bf 79}, 084043 (2009), arXiv: 0902.4669 [gr-qc].
\bibitem{Konno} K. Konno, T. Matsuyama, Y. Asano, S. Tanda, Phys. Rev. D {\bf 78}, 024037 (2008), arXiv: 0807.0679 [gr-qc]; K. Konno, T. Matsuyama, S. Tanda, "Rotating black hole in extended Chern-Simons modified gravity", arXiv: 0902.4767 [gr-qc].
\bibitem{Godel} K. G\"{o}del, Rev. Mod. Phys. {\bf 21}, 447 (1949).
\bibitem{Hawk} S. Hawking, Phys. Rev. D {\bf 46}, 603 (1992).
\bibitem{Reb} M. Reboucas, J. Tiomno, Phys. Rev. D {\bf 28}, 1251 (1983); M. Reboucas, M. Aman, A. F. F. Teixeira, J. Math. Phys. {\bf 27}, 1370 (1985);  M. O. Galvao, M. Reboucas, A. F. F. Teixeira, W. M. Silva, Jr, J. Math. Phys. {\bf 29}, 1127 (1988).
\bibitem{Dabr} M. Dabrowski, J. Garecki, Class. Quant. Grav. {\bf 19}, 1 (2002), gr-qc/0102092.
\bibitem{vayona} J. Barrow, M. Dabrowski, Phys. Rev. D {\bf 58}, 103502 (1998), gr-qc/9803048; P. Kanti, C. E. Vayonakis, Phys. Rev. D {\bf 60}, 103519 (1999), gr-qc/9905032.
\bibitem{Lobo} O. Bertolami, F. Lobo, NeuroQuantol. {\bf 7}, 1 (2009), arXiv: 0902.0559 [gr-qc].
\bibitem{He} X. He, B. Wang, S. Chen, Phys. Rev. D {\bf 79}, 084005 (2009), arXiv: 0811.2322 [gr-qc].
\bibitem{barrow} J. D. Barrow, C. Tsagas, Class. Quant. Grav. {\bf 21}, 1773 (2004), gr-qc/0308067; Phys. Rev. D {\bf 69}, 064007 (2004), gr-qc/0309030; T. Clifton, J. Barrow, Phys. Rev. D {\bf 72}, 123003 (2005), gr-qc/0511076. 
\bibitem{DJPi} S. Deser, R. Jackiw, S.-Y. Pi, Acta Phys, Polon. B {\bf 36}, 27 (2005), gr-qc/0409011; E. Cotton, Ann. Far. Sc. Toulouse (II), {\bf 1}, 385 (1899); in four-dimensional gravity, the Cotton tensor was introduced by R. Arnowitt, S. Deser, C. Misner, gr-qc/0405109.  
\end{thebibliography}
\end{document}